\documentclass[pra,twocolumn,superscriptaddress]{revtex4-1}
\pdfoutput=1
\usepackage{amsmath,graphicx}

\usepackage{xcolor,cancel}

\newcommand{\coeff}{\varphi}
\newcommand{\field}{\hat{\psi}}
\newcommand{\vect}[1]{\mathbf{#1}}

\begin{document}

\title{Bath induced coherence and the secular approximation}
\author{P.~R.~Eastham}
\affiliation{School of Physics and CRANN, Trinity College Dublin, Dublin 2, Ireland.}  

\author{P.~Kirton}
\affiliation{SUPA, School of Physics and Astronomy, University of
  St. Andrews, KY16 9SS, U.K.}

\author{H.~M.~Cammack}
\affiliation{SUPA, School of Physics and Astronomy, University of
  St. Andrews, KY16 9SS, U.K.}

\author{B.~W.~Lovett}
\affiliation{SUPA, School of Physics and Astronomy, University of
  St. Andrews, KY16 9SS, U.K.}

\author{J.~Keeling}
\affiliation{SUPA, School of Physics and Astronomy, University of
  St. Andrews, KY16 9SS, U.K.}

\begin{abstract}
  Finding efficient descriptions of how an environment affects a
  collection of discrete quantum systems would lead to new insights
  into many areas of modern physics. Markovian, or time-local, methods
  work well for individual systems, but for groups a question arises:
  does system-bath or inter-system coupling dominate the dissipative
  dynamics? The answer has profound consequences for the long-time
  quantum correlations within the system. We consider two bosonic
  modes coupled to a bath. By comparing an exact solution against
  different Markovian master equations, we find that a smooth
  crossover of the equations-of-motion between dominant inter-system
  and system-bath coupling exists -- but requires a non-secular master
  equation. We predict a singular behavior of the dynamics, and show
  that the ultimate failure of non-secular equations of motion is
  essentially a failure of the Markov approximation. Our findings
  support the use of time-local theories throughout the crossover
  between system-bath dominated and inter-system-coupling dominated
  dynamics.
\end{abstract}
\maketitle


\section{Introduction}
\label{sec:introduction}

A Markovian system is one in which the future time evolution depends
only on the current state, and not on its history~\cite{vankampen}. In
the context of open quantum systems, Markovianity generally implies
that the reduced density operator obeys a first-order differential
equation. This class of theory has been developed for many years, is
applied to a vast range of systems, and provides our understanding of
quantum damping and decoherence~\cite{Breuer2002}. Recent work,
however, presents it with challenges. The development of solid-state
quantum emitters, such as single and coupled quantum
dots~\cite{Badolato2005,Ramsay2010c} and superconducting-qubit
cavities~\cite{Nissen2013,Henriet2014}, demands theories capable of
treating driven or coupled systems damped by complex structured
baths~\cite{Kaer2010,Roy2011,Kim2014a,Nysteen2013,YugeTatsuroKamideKenji2014}. Such
theories reveal, among other effects, the possibility of engineering
the reservoirs to control quantum
coherence~\cite{Diehl2008,Paavola2010}. They show that under
appropriate conditions both quantum coherence~\cite{Tscherbul2014} and
entanglement~\cite{Paz2008,Benatti2003,McCutcheon2009} can survive
indefinitely, even for high temperature
baths~\cite{Galve2010,Estrada2015}.

These problems do not necessarily elude treatment by a time-local
theory, i.e., a (Markovian) quantum master equation.  Such theories
accurately reproduce the intensity-dependent damping of quantum dots
in a structured reservoir~\cite{Ramsay2010c,Eastham2013,Roy2011a}, for
example, and provide recent predictions of bath-induced
coherence~\cite{Tscherbul2014} and
entanglement~\cite{Benatti2003,McCutcheon2009}. However, there are
several master equations consistent with, and derivable from, the
assumption of weak
coupling~\cite{Benatti2005,Dumcke1979}. Furthermore, master equations
are often postulated phenomenologically, by choice of the jump
operators in the Lindblad form. For problems with multiple oscillators
and structured baths this choice is not straightforward, with
different choices plausible in different limits. Nor is it innocent:
different forms of master equation lead to different
behavior~\cite{Joshi2015}. Thus it is important to establish which, if
any, of the various time-local theories is correct.

In this paper we address this question by studying an exactly-solvable
model, and comparing the exact solution against various time-local
theories.  We consider a model of two bosonic modes, $\field_{a,b}$,
with frequencies $\omega_{a,b}$, coupled to a thermally-occupied bath
with spectral density $J(\nu)$. This model has a non-trivial
Hamiltonian, multiple degrees-of-freedom, and frequency-dependent
damping, yet is exactly solvable. We consider the general case where
the natural frequencies $\omega_{a,b}$ differ and the bath couples to
a superposition of modes, $\coeff_a^\ast \field_a+\coeff_b^\ast
\field_b$, and calculate the evolution of the coherence, $\langle
\field_a^\dagger \field_b\rangle$. We find a complex behavior with
multiple regimes, visible in Fig.~\ref{fig:coherence}, reflecting the competing effects
of the system Hamiltonian and the coupling to the bath. We will
compare the exact solution with time-local theories, and so identify
those which correctly capture such physics. This allows us to
establish their validity in a generic problem, and avoids the
difficulty inherent in studying only approximate theories.

A physical issue we will address is the appropriate form of dissipator
for systems with multiple components.  Two different forms are
expected on physical grounds~\cite{Steck}. In the case of the
two-oscillator model it is clear that at resonance,
$\omega_a=\omega_b$, the damping can depend only on the pattern of
coupling to the baths.  Thus we expect collective decay, described by
a Lindblad form $L_c=\Gamma^\downarrow \mathcal{L}[\coeff_a^{\ast}
  \field_a^{} + \coeff_b^{\ast} \field_b^{}] + \Gamma^\uparrow
\mathcal{L}[\coeff_a\field_a^\dagger + \coeff_b \field_b^\dagger]$,
where $\mathcal{L}[\field]$ is the standard dissipator with jump
operator $\field$~\cite{Breuer2002}. Far off-resonance, however, we
expect individual decay terms, $L_i=\sum_{i=a,b} \Gamma_i^\downarrow
\mathcal{L}[\field_i^{}] + \sum_{i}\Gamma_i^\uparrow
\mathcal{L}[\field_i^\dagger]$. The first form predicts a non-zero
steady-state coherence, while the second predicts this vanishes. We
will show that neither of these forms is, in general, correct, and
both make misleading predictions outside of limiting
cases. Nonetheless, we will find that the general behavior can be
accurately treated by a time-local theory, specifically a
Bloch-Redfield equation. While one might na\"ively have expected some
smooth crossover between the limits captured by $L_c$ and $L_i$, the
real answer is more subtle: a smooth interpolation exists for the
equations-of-motion, but the steady-state is \emph{singular} at
degeneracy. This allows mutually exclusive behavior in different
regimes, and implies that some useful effects -- specifically the
protection of coherence against the bath -- are critically sensitive to
microscopic parameters. Moreover, while the crossover can be treated
by a time-local theory, this theory is not a Lindblad form with the
required positive rates. The use of such forms is the subject of
ongoing debate, since they are not completely positive maps
\cite{Benatti2009}. This means that they can lead to unphysical
density operators, with negative eigenvalues.

A methodological issue in this debate is the procedure of
secularization. This amounts to removing from the equations-of-motion
those terms which are time dependent in the interaction picture. It
was used in some of the earliest work on quantum damping by Bloch and
Wangsness~\cite{Wangsness1953}, but was then argued to be unnecessary
by Redfield~\cite{Redfield1957} as well as
Bloch~\cite{Bloch1957}. That position was challenged by the subsequent
Lindblad theorem~\cite{Lindblad1976}: as argued by D\"umcke and
Spohn~\cite{Dumcke1979}, secularization is required to reach a
description where Lindblad's theorem ensures positivity of the density
operator. Indeed, Lindblad's theorem guarantees that the density
operator will remain positive even when there is entanglement with an
auxiliary system, a criterion known as complete
positivity~\cite{Breuer2002}. Secularization, which leads to a
completely positive theory, is clearly appropriate when the
interaction-picture time dependence is fast, since off-diagonal terms
then rapidly average to zero. In our case, this is far off-resonance,
and secularization indeed leads from the Bloch-Redfield equation to
the form $L_i$. However, for a tunable system it may occur that the
the time-dependence in the interaction picture becomes slow in certain
regimes, i.e., approaching resonance, so that secularization becomes
inappropriate. An interesting improved version of the secularization
procedure is studied in Ref.~\cite{Benatti2009}.

Recently, the necessity of secularization has been
questioned~\cite{Rivas2010,Jeske2014a}: Simulations indicate that for
time evolution following an initially prepared separable state,
secularization (and even a Lindblad form for the equation of motion)
can be unnecessary for positivity~\cite{Whitney2008}, and even
complete positivity~\cite{Maniscalco2007}, in particular for
time-convolutionless~\cite{Breuer2002} and
Nakajima-Zwanzig~\cite{Nakajima1958,Zwanzig1960} approaches.  Stronger
statements to this effect have also been made by Hell \emph{et
  al.}~\cite{Hell2014}, noting that a conservation
law~\cite{Salmilehto2012} is violated by secularized theories --- we
discuss this sum rule in detail further below.  The question of how
the operator form of time-local and non-Markovian approaches are
related is reviewed by Karlewski and Marthaler~\cite{Karlewski2014}.
Our focus in this paper is, however, on cases where a time-local
description is sufficient. This will enable us to explore the entire
parameter space of a model, and identify the regions where
Bloch-Redfield equations predict physical behavior.  We
will show that, although the damping is not of Lindblad form, the
anticipated unstable behavior does not occur within the domain of
applicability of the theory -- specifically, so long as the bath
remains Markovian. 

The exact results we present are restricted to only a subset of
possible initial 
density matrices. We take as initial conditions a thermal state of the
bath and the ground state of the two oscillators. The reduced density
matrix is then Gaussian at all times, and so completely characterized
by its second moments. Thus we will be able to establish whether the
Bloch-Redfield equations are accurate and physical from the dynamics
of those moments alone. This does not, however, rule out inaccurate or
even unphysical behavior for arbitrary (non-Gaussian) initial density
matrices.

Within the scope of coupled open quantum systems, a particular
motivation for our work comes from the timely theory of ``weak
lasing''~\cite{Aleiner2011} introduced in the context of polariton
condensates.  The idea presented is that for modes which are close to
resonance the (dissipative) radiative coupling can select which linear
combination of modes lases (condenses) first.  These works started
from a phenomenological description of radiative coupling, in which
collective dissipation terms are introduced by hand.  In the following
we will see, however, that the effects of collective dissipation terms
are strongly dependent on whether the individual modes are degenerate
or not. Our work does not consider the general problem with both drive
and dissipation, but the results we present for coupling to a single
bath suggest there may be a need to re-examine how weak-lasing evolves
where radiative coupling selects superpositions of non-degenerate
modes.

The remainder of this paper is structured as follows. In
Sec.~\ref{sec:model} we describe the model. In
Sec.~\ref{sec:exact-solution} we present the exact solution, and
discuss its behavior. In Sec.~\ref{sec:bloch-redf-appr} we discuss the
comparison with the Bloch-Redfield equation and the na\"ive Lindblad
forms mentioned above. We also identify the parameter regimes where
the Bloch-Redfield equation gives physical behavior. In
Sec.~\ref{sec:schr-pict-bloch} we develop an alternative to the
Bloch-Redfield equation, and show it to be an improvement both
numerically and analytically. In Sec.~\ref{sec:extens-mult-baths} we
give the generalization of our work to the case of multiple
baths. Finally, in Sec.~\ref{sec:conclusions}, we give our conclusions.

\section{Model}
\label{sec:model}

The two bosons and the common bosonic bath are represented by the Hamiltonian
$\hat{H}=\hat{H}_S+\hat{H}_{SB}+\hat{H}_B$. The system Hamiltonian is $\hat{H}_S = \omega_a
\field_a^\dagger \field_a^{} + \omega_b \field_b^\dagger \field^{}_b$, in
terms of bosonic annihilation operators $\field_i$. The bath Hamiltonian is $\hat{H}_B=\sum_i
\omega_i \hat{c}^\dagger_i \hat{c}^{}_i$, where $c_i$ annihilates a boson in mode $i$. The system-bath coupling takes the form
\begin{equation}
  \label{eq:1}
  \hat{H}_{SB}=
  (\coeff_a^\ast \field_a^\dagger  + \coeff_b^\ast \field_b^\dagger) 
  \sum_i g_i  \hat{c}_i
  + \text{H.c.}.
\end{equation}
The complex coefficients $\coeff_i$ determine which pattern of system
operators the bath couples to, and $g_i$ captures the overall coupling
to mode $i$.  We will assume the bath has a continuous density of
states parameterized by the spectral density $J(\nu) =\sum_i g_i^2
\delta(\nu-\omega_i)$.

Since this model is a linear system of coupled harmonic oscillators it
is exactly solvable. The exact solution for a single harmonic
oscillator coupled to a bath~\cite{Barnett2015} is
well-known~\cite{Grabert1988}. The extension to the case of two
identical oscillators coupled symmetrically to a bath can be found in
Ref.\ \cite{Chou2008}, and has been used to test master equation
approaches\ \cite{Rivas2010}. In this special case the normal modes
exactly match the pattern of bath coupling. The antisymmetric mode
then decouples from the bath, immediately reducing the problem to one
damped oscillator and one undamped one. The dynamics of entanglement
in this case was studied by Paz and Roncaglia~\cite{Paz2008}, who
showed that the undamped mode allows entanglement to persist
indefinitely. We consider a more general problem, including detuning
$\Delta=(\omega_a-\omega_b)/2\neq 0$, which prevents such a decoupling
and leads to finite lifetimes.

\begin{figure}[htpb]
  \centering
  \hspace{-0.3cm}\includegraphics[width=3.5in]{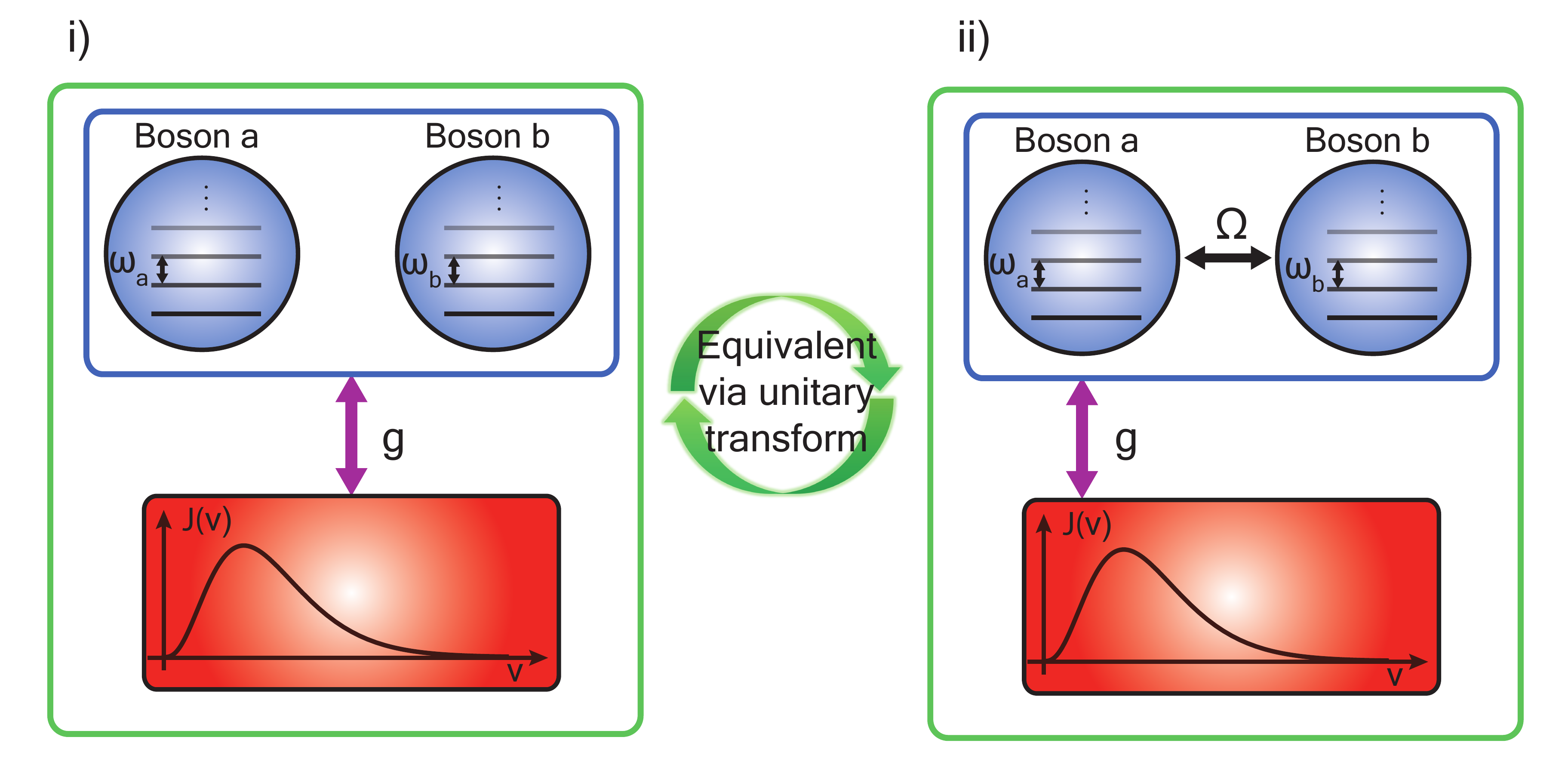}
  \caption{(Color online) Cartoon of the system we consider: (i) two
    bosonic modes of frequencies $\omega_a, \omega_b$ couple
    collectively to a single bath.  As illustrated, we take a
    super-Ohmic bath with an exponential cutoff when an explicit form
    is required.  (ii) the equivalent problem of two
    coupled modes with a bath coupling to only one of the modes.}
  \label{fig:schematic}
\end{figure}

The existence of finite lifetimes at non-zero $\Delta$ can be
understood by observing that the model above is equivalent to a system
of two coupled oscillators, one of which is coupled to a bath, i.e.,
the Hamiltonian $\hat{H}_S = \omega_c \field_c^\dagger \field_c^{} +
\omega_d \field_d^\dagger \field_d^{} + \Omega \field_c^\dagger
\field_d^{} + \text{H.c.}$ with $\hat{H}_{SB} = \field_c^\dagger
\sum_i g_i \hat{c}_i + \text{H.c.}$. This mapping follows on
transforming this latter problem to a basis in which $\hat{H}_S$ is
diagonal.  These two  equivalent problems are illustrated
schematically in Fig.~\ref{fig:schematic}.  We will use the basis of
Eq.~(\ref{eq:1}) in the following; the results of the other problem
can be simply extracted by the appropriate rotations.

In what follows, we consider the time evolution of the observables
$F_{ij}(t) \equiv \langle \field_i^\dagger \field_j^{} \rangle$,
focusing in particular on the coherence $F_{ab}(t)$ which, as
mentioned earlier, distinguishes collective from individual
decay. Furthermore, these observables fully characterize the density
matrix for the initial conditions we consider; it is Gaussian, so that
higher moments are related to the $F_{ij}$ by Wick's theorem. We first
present the exact solution and discuss its observed properties, before
considering the (non-secularized) Bloch-Redfield (BR) equation of
motion.  We will show both analytically and numerically that this
approach reproduces the exact solution, while either of the na\"ive
Lindblad master equations fail to reproduce the exact results.

\section{Exact solution}

\label{sec:exact-solution}
The exact time evolution can be readily found by using a Laplace
transform to write the system operators in terms of the $t=0$ bath
operators, and then evaluating $F_{ij}(t)$ using thermal correlations
for the bath operators at $t=0$. With the oscillators in the ground
state at $t=0$ we find:
\begin{align}
  \label{eq:2}
  F_{ij}(t) &= \int d\nu J(\nu) n_B(\nu)
  W^\ast_i(\nu,t) 
  W^{}_j(\nu,t), 
  \\
  \label{eq:3}
  W^{}_i(\nu,t) &=
  \coeff_i\int \frac{d\zeta}{2\pi} 
  \frac{(\omega_{\bar{i}}-\zeta)e^{-i\zeta t}}{(\nu-\zeta-i0)d(\zeta+i0)},
\end{align}
where $\omega_{\bar{a}}=\omega_b$ and vice versa, $n_B(\nu)$ is the
Bose-Einstein distribution function, and $d(\zeta) =
-(\omega_a-\zeta)(\omega_b-\zeta)
+iK^\ast(\zeta)[|\coeff_a|^2(\omega_b-\zeta)
  +|\coeff_b|^2(\omega_a-\zeta)]$ is the denominator of the retarded
Green's function.  Here we have introduced $K(\zeta)$, the analytic
continuation of the damping rate to the lower half plane $K(\zeta) = i
\int dx J(x)/(x - \zeta + i0)$. For real $\zeta$ the real part of
$K(\zeta)$ is proportional to the spectral density, while the
imaginary part follows from a Kramers-Kronig relation. In the
numerical results which follow we use the form of spectral density
illustrated in Fig.~\ref{fig:schematic}. For numerical evaluation, it
is computationally more efficient to write this as a convolution:
\begin{align}
  \label{eq:4}
  F_{ij}(t) &=  \int_0^t \!d\tau\!\int_0^t \!d\sigma
  D_i(t-\tau)^\ast D_j(t-\sigma) 
  \alpha(\sigma-\tau),
  \\
  \label{eq:5}
  D_i(t)&=
  \!\coeff_i\int \frac{d\zeta}{2\pi} 
  \frac{(\omega_{\bar{i}}-\zeta)e^{-i\zeta t}}{d(\zeta+i0)}, 
  \\
  \label{eq:6}
  \alpha(\tau)&=\int\! d\nu J(\nu) n_B(\nu) e^{-i \nu \tau}.
\end{align}
One may readily check that this is equivalent to
Eqs.~(\ref{eq:2},\ref{eq:3}).


\subsection{Behavior near degeneracy}
\label{sec:behav-near-degen}

\begin{figure}[htpb]
  \centering
  \includegraphics[width=3.2in]{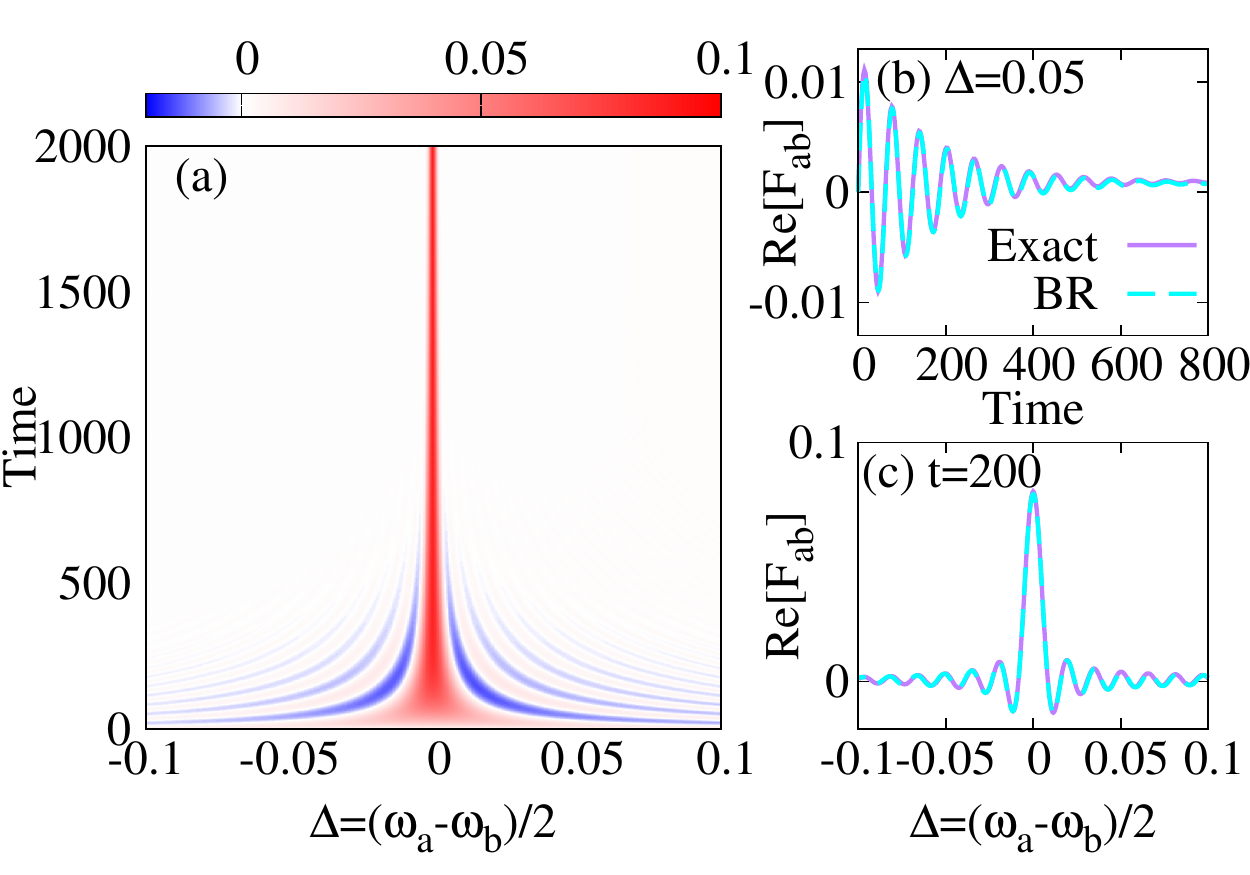}
  \caption{(Color online) (a) Evolution of coherence $F_{ab}$ with time (vertical)
    and detuning (horizontal), found from the exact solution. We use a
    super-Ohmic density of states with an exponential cutoff,
    $J(\omega)=J_0 e^{z - z \omega/\omega_0}(\omega/\omega_0)^z$.
    This form is written such that its peak value is at
    $\omega=\omega_0$, and $J(\omega_0)=J_0$.  We choose $z=3$, and we
    measure all energies and times in units such that $\omega_a=1$.
    In these units $\omega_0=0.9, J_0=0.001$, and the thermal
    occupation is controlled by $k_BT=0.52$. The horizontal axis is
    the detuning, found by varying $\omega_b$ for fixed $\omega_a$.
    (b) Vertical slice at $\omega_b=0.9$ corresponding to
    $\Delta=0.05$, showing comparison between exact and Bloch-Redfield
    theories. (c) Horizontal slice at $t=200$. In the secularized
    theory the coherence $F_{ab}=0$ for all times.}
  \label{fig:coherence}
\end{figure}

Using the above expressions, one may directly find how the coherence
evolves with time, as the detuning $\Delta=(\omega_a-\omega_b)/2$
changes; this is shown in Fig.~\ref{fig:coherence}. As is clear in
this figure, the behavior at degeneracy ($\Delta=0$) and away from
degeneracy is different. Near degeneracy there is strong,
  long-lived coherence, corresponding to the noise-induced coherence
  recently analyzed for few-level
  models~\cite{Tscherbul2014,McCutcheon2009}. What is not immediately
clear however is that the degenerate limit is in fact singular: the
form of $F_{ab}(\infty)$ is discontinuous as a function of frequency.
We next turn to discuss how and why this singular behavior occurs.

\newcommand{\pole}{\zeta^{0}} The origin of the discontinuity at
degeneracy is the emergence of a slow mode, whose lifetime diverges as
$\Delta \to 0$. The decay rates of oscillations can be extracted from
the poles $\pole$ of the the Keldysh Green's function, i.e.\ solutions
of $d(\pole)=0$.  Writing $\omega_{a,b} = \Omega \pm \Delta$ leads to
an expression:
\begin{multline}
  \label{eq:7}
  0 = \Delta^2 + \Delta (|\coeff_a|^2 - |\coeff_b|^2) iK^\ast(\pole)
  \\
  -(\Omega-\pole)
  \left(\Omega-\pole-iK^\ast(\pole) [|\coeff_a|^2 + |\coeff_b|^2] \right).
\end{multline}
At $\Delta=0$, the first line vanishes so it is clear that there is a
pole $\pole=\Omega$, which is real and so entirely undamped.  This has
a simple physical interpretation: at $\Delta=0$, there is no coupling
between the combination $\sum_i \coeff_i \field_i$ and the orthogonal
combination of fields.  As such, the orthogonal combination is
entirely undamped, and maintains its original state.  With non-zero
detuning, beating between the modes $\field_{i=a,b}$ means the
orthogonal combination evolves into $\sum_i \coeff_i \field_i$ with
time, and is thus damped.  Considering small $\Delta$ perturbatively
gives:
\begin{equation}
  \label{eq:8}
  \Im[\pole] = 
  -\frac{%
    4  \Delta^2 |\coeff_a|^2 |\coeff_b|^2}{(|\coeff_a|^2 + |\coeff_b|^2)^3}
  \frac{K^\prime(\Omega)}{|K(\Omega)|^2}
  + \mathcal{O}(\Delta^3).
\end{equation}
This explains the Gaussian form of the singular response visible in
Fig.~\ref{fig:coherence}: we expect $F_{ab}(\Delta,t\to \infty) \sim
\bar{F}_{ab}(\Delta) + C\exp( - \alpha \Delta^2 t)$ at large $t$, where
$C,\alpha$ are constant factors and $\bar{F}_{ab}(\Delta)$ is a smooth
function.

\subsection{Late time asymptotes}
\label{sec:late-time-asymptotes}

The long-time asymptotes of the observables can be obtained from the
pole structure of the Laplace-transform solution \cite{Stace2013}. For
late times, $W_i(\nu,t)$ simplifies significantly because the pole at
$\zeta=\nu-i0$ lies on the real axis and so has a vanishing decay
rate, while the poles of $d(\zeta+i0)$ are generically off axis and so
have decayed at late times.  Thus $W_i(\nu, t \to \infty) =
[-ie^{-i\nu t}] \coeff_i(\omega_{\bar{i}}-\nu)/d(\nu+i0)$. This gives
a simplified expression
\begin{equation}
  \label{eq:9}
  F_{ij}(\infty)
  =
  \coeff^\ast_i \coeff^{}_j
  \int d\nu J(\nu) n_B(\nu)
  \frac{(\omega_{\bar{i}}-\nu)(\omega_{\bar{j}}-\nu)}{|d(\nu+i0)|^2}.
\end{equation}
As can just be seen in Fig.~\ref{fig:coherence}, away from the
resonance point, the off-diagonal coherence decays at late times to a
small value, but not strictly to zero.  However, if the bath density
of states and occupation are strictly flat, i.e.\ if $J(\nu)=J_0,
n_B(\nu)=n_0$, then one may show that the asymptotic value
$F_{ij}(\infty)$ vanishes for $\Delta \neq 0$.  In this case
Eq.~(\ref{eq:9}) simplifies considerably, as $K(\nu)=\pi J_0$ for a
flat bath, so $d(\nu+i0)$ becomes a simple polynomial.  This integral
then has only four simple poles, and one may readily check that it
exactly vanishes -- except at $\omega_a=\omega_b$, where two of the
poles coincide and cancel with the zeros of the numerator.  The small
residual coherence that exists away from resonance in
Fig.~\ref{fig:coherence} is thus due to the frequency dependence of
$n_B(\nu), J(\nu)$.


\section{Bloch-Redfield approach}
\label{sec:bloch-redf-appr}

So far, we have seen that the exact solution of the bosonic problem
does show a crossover between strong coherence at degeneracy and weak
coherence, due to a frequency-dependent spectral density, away from
degeneracy. However, this crossover occurs as a function of time, with
coherence surviving over a range $\alpha t \simeq \Delta^{-2}$. A
similar quadratically diverging timescale is found in the V-type
system~\cite{Tscherbul2014}. We now turn to consider whether the
behavior of the coherence, and other observables, can be reproduced by
a time-local master equation.

We may first note that neither of the na\"ive forms (individual or
collective damping) discussed in the introduction reproduce the
correct behavior. Separate decay predicts that the coherence vanishes,
for all times and detunings. The collective decay model does predict a
strong long-lived coherence close to resonance, and indeed a
quadratically-diverging lifetime.  After this time, however, the
coherence decays to zero, rather than the non-zero value predicted by
the exact solution. More significantly, however, the collective form
fails to reproduce the behavior once the detuning becomes
significant. This may be seen from the steady-state populations: for
large detuning or weak-coupling these correspond to equilibrium with
the bath, so $F_{aa}=n_B(\omega_a), F_{bb}=n_B(\omega_b)$, whereas the
Lindblad form gives equal populations. As pointed out by Cresser for
the Jaynes-Cummings model~\cite{Cresser1992c}, such master equations
do not reach canonical equilibrium. More generally, since $L_c$ is
parameterized by one pair of forward/backward rates, it cannot account
for the presence of two frequencies in the dynamics at which the bath
should be sampled. As can be seen from Fig. \ref{fig:coherence}, this
occurs above a critical value of the detuning. Thus this model cannot
possibly be accurate in this regime, unless the bath and its occupation are flat.

Thus, neither na\"ive form of dissipator can give a full account of problems with multiple
system frequencies and structured baths, particularly if one seeks to
analyze coherence. Unfortunately many interesting problems in
solid-state quantum optics fall in this class, as discussed in the
introduction.  We will now show, however, that a Bloch-Redfield
equation does reproduce the correct behavior, as long as one does not
secularize the final result.  Such an approach is frequently stated to
be invalid, as it leads to negative rates and instabilities.  We will
however show analytically that such instabilities occur in a much
restricted parameter regime, and, in fact, only when the Markov
approximation breaks down. The non-secularized theory is, also, often
argued to be invalid on the related grounds that it is not a
completely positive map, and may not even be a positive one. We will
however show analytically that, although the map is not positive, it
preserves positivity for almost all Gaussian states. Furthermore, we
find numerically that these states soon dominate under the time
evolution, even if dangerous ones are present in the initial
conditions.

Following the standard method~\cite{Breuer2002} one finds the 
master equation has the form:
\begin{multline}
  \label{eq:10}
    \partial_t \rho = -i [\hat H, \rho] 
  + \sum_{ij}  {L}_{ij}^\downarrow \coeff_i^\ast \coeff_j^{}
  \left(2 \field^{}_j \rho \field^\dagger_i
    - [\rho, \field^\dagger_i \field^{}_j]_+ \right)
  \\+ \sum_{ij}  {L}_{ij}^\uparrow  \coeff_i^{} \coeff_j^\ast
  \left(2 \field^\dagger_j \rho \field^{}_i
    - [\rho, \field^{}_i \field^\dagger_j]_+ \right).
\end{multline}
Here the Hamiltonian includes Lamb shifts $\hat{H}=\hat{H}_S - \sum_{ij} 
{h}_{ij}  \coeff_i^\ast \coeff_j \field^\dagger_i \field^{}_j$. The 
matrices $L^{\sigma \in \downarrow,\uparrow},h$ can be written in a compact form,
\begin{align}
  \label{eq:11}
  L^{\sigma}&=
  \begin{pmatrix}
    K_{a\sigma}^{\prime}
    &
    \bar{K}_{\sigma}^{\prime}
    \pm 
    i \delta{K}_{\sigma}^{\prime\prime}
    \\
    \bar{K}_{\sigma}^{\prime}
    \mp
    i \delta{K}_{\sigma}^{\prime\prime}
    &
    K_{b,\sigma}^{\prime}
  \end{pmatrix}
  \\
  \label{eq:12}
  h&=
  \begin{pmatrix}
    K_a^{\prime\prime}
    &
    \bar{K}^{\prime\prime} - i \delta K_{}^{\prime} 
    \\
    \bar{K}^{\prime\prime} + i \delta K_{}^{\prime} 
    &
    K_b^{\prime\prime}
  \end{pmatrix},
\end{align}
with the upper (lower) signs in Eq. (\ref{eq:11}) for $L^{\downarrow}$
($L^{\uparrow}$). Here we have introduced several new pieces of
notation. We have used the shorthand $K_i=K(\omega_i)$ in terms of the
Hilbert transform (analytic continuation) defined previously, and have
also defined Hilbert transforms of the excitation (absorption) rate
$K_{i\uparrow}= i\int d\xi n_B(\xi) J(\xi)/(\xi - \omega_i +i0)$, and
de-excitation (emission) rate $K_{i\downarrow}= i\int d\xi
(n_B(\xi)+1) J(\xi)/(\xi - \omega_i +i0)$.  Note that this means $K_i
= K_{i\downarrow} - K_{i\uparrow}$.  Primes signify real and imaginary
parts and $\bar{X} = (X_a + X_b)/2, \delta{X}=(X_a - X_b)/2$.

While Eqs.~(\ref{eq:10}--\ref{eq:12}) fully describe the equations of
motion, it is more convenient to use the (closed) set of equations of
motion for the quantities $F_{ij}^{}$ derived from these master
equations.  In order to simplify these equations, it is convenient to
note that the phase of the complex coefficients $\coeff_i$ can be
eliminated by a phase twist of the original operators, and we thus assume
$\coeff_i$ is real from hereon.  We may then define the vector of real
quantities $\vect{f}=(F_{aa}^{}, F_{bb}^{}, 2F_{ab}^\prime,
2F_{ab}^{\prime\prime})^T$ and produce an equation of motion
$\partial_t \vect{f} = -\vect{M} \vect{f} + \vect{f}_0$ where 
\begin{equation}
  \label{eq:13}
  \vect{M}
  =
  \begin{pmatrix}
    2\coeff_a^2 K^{\prime}_a  & 0 &
    \coeff_a \coeff^{}_b K^{\prime}_b &
    \coeff_a \coeff^{}_b K^{\prime\prime}_b 
    \\
    0 & 2\coeff_b^2 K^{\prime}_b  &
    \coeff_b \coeff^{}_a K^{\prime}_a &
    -\coeff_b \coeff^{}_a K^{\prime\prime}_a
    \\
    2 \coeff_a \coeff^{}_b K^{\prime}_a
    &
    2 \coeff_b \coeff^{}_a K^{\prime}_b
    &  \Gamma_0 & -E_0 
    \\
    -2 \coeff_a \coeff^{}_b K^{\prime\prime}_a
    &
    2 \coeff_b \coeff^{}_a K^{\prime\prime}_b
    & E_0 &  \Gamma_0 
  \end{pmatrix},
\end{equation}
with $E_0 = (\omega_b-\coeff_b^2 K^{\prime\prime}_b)
- (\omega_a - \coeff_a^2 K^{\prime\prime}_a)$, and
$\Gamma_0=\coeff_a^2 K^{\prime}_a + \coeff_b^2 K^{\prime}_b$.
None of these rates depend on the bath mode occupations, however the constant vector 
\begin{math}
\vect{f}_0=2(
\coeff_a^{2}K_{a,\uparrow}^{\prime},\ 
\coeff_b^{2}K_{b,\uparrow}^{\prime},\ 
\coeff_a \coeff_b 2 \bar{K}_{\uparrow}^{\prime},\ 
-\coeff_a \coeff_b 2\delta K_{\uparrow}^{\prime\prime})^T
\end{math}
 involves the excitation rate, so that populations are
proportional to the bath occupations as expected.

The result of time evolving this closed set of equations is shown in
Fig.~\ref{fig:coherence}(b,c), and clearly compares very well to the
exact solution.  Moreover, we can easily see that secularizing this
set of equations, as is often claimed to be a crucial
step~\cite{Dumcke1979}, could only decrease the agreement:
secularization can be shown to be equivalent to setting all terms
involving the product $\coeff_a\coeff_b$ to zero, thus removing the
off-diagonal blocks of Eq.~(\ref{eq:13}) and the last two elements of
the vector $\vect{f}_0$.  This then makes the coherence $F_{ab}(t)$
identically zero.  This is as expected for a secular theory: a
non-zero detuning $\omega_a\neq\omega_b$ means the master equation
contains no cross terms between modes $a,b$ and thus no coherence
arises. Note that the coherence in the secular theory is identically
zero, whereas that in the exact result decays to a small value after
the time $1/(\alpha \Delta^2)$, see Eq. (\ref{eq:8}). We can thus
identify this timescale as that controlling the secular approximation.

\subsection{Stability of time evolution}
\label{sec:stab-time-evol}

The frequently stated reason~\cite{Dumcke1979} for secularizing the
equation of motion is that it is required to ensure the equation is of
Lindblad form with positive rates, i.e.\ that the master equation take
the form $\dot{\rho} = -i[H,\rho] +\sum_i \lambda_i
(2\hat{\Lambda}^{}_i \rho \hat{\Lambda}^\dagger_i
-[\hat{\Lambda}^\dagger_i\hat{\Lambda}^{}_i,\rho]_+)$ with
$\lambda_i\ge 0$.  This is desired so that Lindblad's theorem can
guarantee complete positivity of the density matrix.  In addition, negative
decay rates may lead to exponentially growing observables.  Despite
its near-perfect match to the exact solution, our non-secularized
equation clearly fails these requirements.  Eq.~(\ref{eq:10}) can be
put into Lindblad form by diagonalizing the matrices
$L^{\sigma\in\uparrow,\downarrow}$ in Eq.~(\ref{eq:11}), however the
eigenvalues are $\lambda_\sigma = \bar{K}_\sigma^{\prime} \pm
S_\sigma$ where $S_\sigma^2 = (\bar{K}_\sigma^{\prime})^2 +
|\delta{K}_\sigma^{}|^2 \ge (\bar{K}_\sigma^{\prime})^2$.  This means
that except when $\delta K_\sigma =0$, one rate is always negative.
Despite this, there have been several recent works~\cite{Jeske2014a}
which suggest it is not established that this formal problem leads to
any practical difficulties in applying such a theory.

In our problem, we are able to find precise conditions under which the
negative rates in the Lindblad form cause a practical problem.
Operationally, our problem is to solve the four linear coupled
equation for the components $F_{ij}$.  This method will fail if the
matrix $\vect{M}$ has negative eigenvalues.  For a Gaussian problem
such as the one we consider here this condition is in fact the only
practical consideration; all higher moments factorize by Wick's
theorem and so positivity of the eigenvalues of $\vect{M}$ ensures the
dynamics remains bounded.  Remarkably, the eigenvalues of $\vect{M}$
can be found in closed form. They are $\vect{M} \phi_i = \mu_i \phi_i$
with
\begin{align}
  \label{eq:14}
  \mu_i&=\Re[\tilde{K}_a +\tilde{K}_b]
  \pm \sqrt{\Re[Q] \pm |Q|},\\ 
  Q &= 2  \tilde{K}_a  \tilde{K}_b
  + \frac{1}{2} \left[
    \tilde{K}_a  -\tilde{K}_b
    + i (\omega_a - \omega_b)
  \right]^2 \nonumber
\end{align}
where $\tilde{K}_i = \coeff_i^2 K_i$.  It is clear that when
$\omega_a=\omega_b$, one finds $Q= [\tilde{K}_a +\tilde{K}_b]^2/2$,
which means $\Re[Q] + |Q| = (\Re[\tilde{K}_a +\tilde{K}_b])^2$.  Thus the 
Bloch-Redfield form recovers the fact there is a zero
eigenvalue at degeneracy.

    From this closed form we may check that the eigenvalues $\mu_i$
    remain positive (stable) as long as
\begin{equation}
  2 \Delta^2 \tilde{K}_a^\prime \tilde{K}_b^\prime
  + 
  \Delta(\tilde{K}_a^\prime +\tilde{K}_b^\prime)
  (\tilde{K}_a^\prime \tilde{K}_b^{\prime\prime}
  -
  \tilde{K}_b^\prime \tilde{K}_a^{\prime\prime}) > 0.
  \label{eq:15}
\end{equation}
The first term is always positive, and thus instability require two
conditions: Firstly, it requires that $\Delta (\tilde{K}_a^\prime
\tilde{K}_b^{\prime\prime} - \tilde{K}_b^\prime
\tilde{K}_a^{\prime\prime}) < 0$, placing a constraint on the
frequency dependence of $J(\nu)$ --- typically an instability is hard
to achieve if $J(\nu)$ has only a single peak, but is possible for a
multi-peaked structure.  Secondly, and more importantly, in order for
the second term in Eq.~(\ref{eq:15}) to dominate, $d K(\omega)/d
\omega$ must be large enough --- this corresponds directly to
requiring that the spectral density should vary significantly on a
scale $J(\omega)$, i.e.\ that the memory time of the bath is
comparable to the damping timescale.  If such a condition is
satisfied, then the Markov approximation is \textit{a priori} invalid.

To summarize, as long as the Markov approximation
is valid \textit{a priori} -- i.e.\ the bath memory time is short compared to damping time --
then the eigenvalues of $\vect{M}$ are positive
and the solution is stable.  This result shows that
Markovianity is, for this problem, a sufficient condition
for stability.
This is despite the Lindblad matrices
$L^{\uparrow,\downarrow}$ always
having negative eigenvalues, except at resonance.  

\subsection{Comparison to exact solution near degeneracy}
\label{sec:comp-exact-solut}

We have already seen the numerical agreement between this BR treatment
and the exact result in Fig.~\ref{fig:coherence}.  We may note that
near resonance one can compare the perturbative solution of the exact
problem to a perturbative expansion of the BR eigenvalues.  Starting
from Eq.~(\ref{eq:14}), and expanding up to quadratic order in
$\Delta$ and $\delta K$, one finds:
\begin{multline}
  \label{eq:16}
  \mu_0 
  =
  \frac{8\coeff_a^2 \coeff_b^2 \Delta^2 \bar{K}^\prime}{(\coeff_a^2 + \coeff_b^2)^3 |\bar{K}|^2}
  \\-
  \frac{8\coeff_a^2 \coeff_b^2 \Delta
    (\bar{K}^\prime \delta K^{\prime\prime} -\delta{K}^\prime \bar{K}^{\prime\prime})
  }{(\coeff_a^2 + \coeff_b^2)^2 |\bar{K}|^2}
  + \mathcal{O}(\Delta^3).
\end{multline} 
Recall that $\delta K$ depends on the detuning, vanishing at least linearly as $\Delta\rightarrow 0$, so that the second term is at least second-order in $\Delta$.
This eigenvalue can be compared to the exact perturbative result by referring back to Eq.~\ref{eq:8} and 
noting that $\mu^{\text{exact}}_0 = -2\Im[\pole]$. The factor of two
appearing here is because $\mu$ corresponds to the eigenvalue of the
population equation, whereas the pole in Eq.~\ref{eq:8} gives the decay 
of fields $\field_i$.

Comparing Eq.~(\ref{eq:8}) to
Eq.~(\ref{eq:16}) one sees that the leading-order term in $K(\omega)$
is correct, but the second term in Eq.~(\ref{eq:16}) is not there in the exact
solution.  The second term is however dependent on the derivative of
the function $K$. Thus one finds again that the BR theory is correct
\emph{as long as the Markovian approximation holds}, i.e.\ as long as
the derivative of the density of states is sufficiently small.

\subsection{Positivity of time evolution}

As we have seen, the Bloch-Redfield time evolution is stable, and has
the correct steady-state, so long as the Markovian approximation is
justified. This rules out the most dramatic pathologies that could
arise from the negative rates, and suggests that the dynamics will not
stray far from the correct behavior. This is consistent with the
essentially perfect agreement seen numerically. We now consider a
related issue, of the extent to which the negative rates lead to
unphysical density matrices with negative eigenvalues.

We first summarize some standard definitions~\cite{Breuer2002}. An
operator is positive if all its expectation values are positive, and a
map is positive if it is between positive operators. Since density
operators are positive the exact time-evolution superoperator, which
is a map between density matrices, is positive. The secularized master
equation in fact satisfies the stronger criterion of complete
positivity, which corresponds to positivity in the presence of
arbitrary entanglement with an auxiliary system.

The map given by Eq.~(\ref{eq:10}) can be shown to be
  non-positive specifically because of the negative eigenvalues of
the Kossakowski matrices $L^{\uparrow,\downarrow}$. To demonstrate
this we suppose that $L^{\downarrow}$ has a negative eigenvalue, and
work in its diagonal basis. We denote the field operator corresponding
to the unstable (stable) eigenvector by $\field_{c}$ ($\field_{d}$),
so that there will be terms in Eq.~(\ref{eq:10}) of the
form \begin{equation} r \left(2 \field_c \rho
  \field_c^\dagger-[\rho,\field_c^\dagger
    \field_c]_+\right)\label{eq:unstjumpterm}\end{equation} with
$r<0$. In general neither $h_{ij}$ nor $L^{\uparrow}$ will be diagonal
in this eigenbasis of $L^{\downarrow}$, so that $H$ contains
terms $\field_i^\dagger \field_j$, for all pairs $i,j \in
c,d$. Similarly, we will have terms in Eq. (\ref{eq:10}) from
$L^{\uparrow}$ of the form $2\field_j^\dagger \rho
\field_i-[\rho,\field_i\field_j^\dagger]_+$, for all such
pairs. However, as positivity requires that all positive
  operators are mapped to positive operators, showing it is violated
  only requires us to construct a single counterexample of a positive
  operator mapped to a non-positive operator, and it is possible to do
  this despite the non-diagonal nature of these other terms. To
  construct this counterexample we suppose $\rho$ describes a pure
Fock state in the diagonal basis of $L^{\downarrow}$,
$\rho=|n,m\rangle\langle n,m|$. This is a positive operator, which is
mapped by the first term in Eq. (\ref{eq:unstjumpterm}) to
$2rn|n-1,m\rangle\langle n-1,m|$. Furthermore, we see that no other
term in the infinitesimal time-evolution superoperator $\Phi(\rho)$
generates this operator. The Hamiltonian and anticommutator terms in
Eq. (\ref{eq:10}) conserve the total excitation number, while the jump
terms from $L^{\uparrow}$ increase it. Thus $\langle
n-1,m|\Phi(\rho)|n-1,m\rangle = 2rn <0$.  Since a positive
  operator $X$ obeys $\forall {| \psi\rangle}: \langle \psi |X |\psi
  \rangle >0$, this fact proves that the map has taken a positive
operator to a non-positive operator. This proves that the map is not
positive. It follows immediately that it is not completely
positive. An analogous argument applies for a negative rate in
$L^{\uparrow}$.

While the Bloch-Redfield Eq. (\ref{eq:10}) is not positive, it
nonetheless agrees well with the exact solution. This suggests that
the operators which are mapped out of the physical space, such as the
one constructed above, are absent from, or at least a negligible
contribution to, the density matrix. To investigate this, and explore
the domain of validity of the theory more generally, we consider
whether the dynamics is positive for Gaussian states. We consider
specifically the subset of Gaussian states relevant to the dynamics
above, where the baths and initial conditions are such that
$G_{ij}=\langle \field_i \field_j\rangle=0$.

For Gaussian states the density matrix is positive if the uncertainty
principle is satisfied~\cite{Simon2000}, which here is equivalent to
$F_{ij}$ being positive semi-definite. This follows on noting that for
two oscillators any normalized linear combination of the operators
$\field_a,\field_b$ is a lowering operator $\hat\eta$, with
corresponding quadratures
$\hat{x}=(\hat\eta+\hat\eta^\dagger)/\sqrt{2},
\hat{p}=-i(\hat\eta-\hat\eta^\dagger)/\sqrt{2}$, and requiring $\Delta
x\Delta p \geq 1/2$ for all such quadratures. Positivity of the
density matrix can thus be checked numerically by calculating the
smallest eigenvalue of $F_{ij}$. In the Bloch-Redfield solution
corresponding to Fig. \ref{fig:coherence}(a) we find that there is a
brief transient period, up to $t\approx 1$, where the state violates
positivity by a tiny amount. Specifically, the smallest eigenvalue of
$F_{ij}$ reaches $\lambda_{m}\sim -10^{-4}$ in this regime, after
which it it is always positive or zero, with typical values
$\lambda_{m}\sim 0.1$. More generally, the Bloch-Redfield $\lambda_m$
agrees with the exact result to four decimal places. The error is
hardly noticeable, except in that it takes the results slightly
outside the physical regime at early times.

The behavior discussed above can be understood by deriving the condition under which
the Bloch-Redfield Eq.  (\ref{eq:10}) preserves positivity for
Gaussian states. For a time increment $\Delta t$ the Bloch-Redfield
Eq. (\ref{eq:10}) implies a shift in the $F_{ij}$, $F_{ij}\rightarrow
F_{ij}^{(0)}+ \Delta t R_{ij}$. Since the time evolution of the
density matrix is continuous it can only become unphysical if
$F_{ij}^{(0)}$ has a zero eigenvalue, which becomes negative under the
perturbation $R_{ij}$. Such an $F_{ij}^{(0)}$ must be of the
form \begin{equation}\begin{pmatrix} n_a & \sqrt{n_a n_b}e^{i\phi} \\
    \sqrt{n_a n_b}e^{-i\phi} &
    n_b\end{pmatrix},\label{eq:gscovmat}\end{equation} with
$n_a,n_b\geq 0$. From the forms of $\vect{M}$ and $\vect{f}_0$ we
calculate the shift matrix elements $R_{ij}$ for the state
$F_{ij}^{(0)}$. We can then calculate the shift in the zero eigenvalue
perturbatively, and find it to be negative
when \begin{multline}\label{eq:posvcrit}\coeff_a^2
  K_{a,\uparrow}^\prime n_b + \coeff_b^2 K_{b,\uparrow}^\prime n_a \\
  - 2\coeff_a\coeff_b\sqrt{n_a n_b}[\bar{K}^\prime_\uparrow
  \cos(\phi)-\delta K^{\prime\prime}_{\uparrow}\sin(\phi)] <
  0.\end{multline} This condition gives a range of $n_a-n_b$ and
$\phi$ for which the minimum-uncertainty Gaussian state,
Eq. (\ref{eq:gscovmat}), is mapped out of the space of physical
states. If the rates are not too different, i.e., the Markov
approximation is well satisfied, then this range is small.

In summary, the two-mode Bloch-Redfield equation is positive for most
Gaussian states. The exceptions are rare, being the subset of
minimum-uncertainty states defined by Eq. (\ref{eq:posvcrit}). Since
the dissipation drives the system towards safe Gaussian states these
dominate the dynamics, even if the others are present in the initial
conditions. Indeed, a positivity-violating state is present in the
initial condition for Fig. \ref{fig:coherence}, but its effects are
transient and quantitatively small.

\section{Beyond the Bloch-Redfield equation}
\label{sec:schr-pict-bloch}


From the above we may conclude that over a wide range of parameters
the Bloch-Redfield theory without secularization accurately matches
the exact solution, while secularization reduces the accuracy.  This
however leaves open an alternate question: does the BR master
equation, and the corresponding coupled equations of motion for
$F_{ij}(t)$, represent the best possible time-local theory of this
problem?  In this section we show that a better set of time-local
equations exists, and involves a minor change to the form of the
matrix $\vect{M}$ that appears in Eq.~(\ref{eq:13}).

\subsection{Sum rule violation}
\label{sec:sum-rule-violation}

There are two motivations to suggest that an improved equation is
possible.  The first is that, as noted above, the BR prediction for
the slowest decay rate of coherence, Eq.~(\ref{eq:16}), does not match
the rate derived from the poles of the exact solution,
Eq.~(\ref{eq:8}).  The second reason concerns sum rules as discussed
in~\cite{Salmilehto2012,Hell2014}. These state that for operators
which commute with the system-bath coupling the time evolution of
such operators in the full dynamics should be equal to that in the
absence of system-bath coupling. For our model, the operators $\hat{X}
= \coeff^\ast_b \field_a - \coeff^\ast_a \field_b$ and
$\hat{X}^\dagger$ obviously commute with Eq.~(\ref{eq:1}).  As such,
their time derivatives should be the same as that following from
$\hat{H}_S$ alone.  In terms of population equations this corresponds
to the statement that
\begin{displaymath}
  I \equiv \langle \hat{X}^\dagger \hat{X} \rangle
  = |\coeff_b|^2 F_{aa} + |\coeff_a|^2 F_{bb}
  - 2 \Re[ \coeff_a^\ast \coeff_b F_{ab} ]
\end{displaymath}
should obey $\partial_t I = \Re[2 i (\omega_a-\omega_b) \coeff_a^\ast
\coeff_b F_{ab} ]$.  In the case that $\coeff_i$ are real, this means that
one should have:
\begin{equation}
  \label{eq:19}
  \begin{pmatrix}
    \coeff_b^2 \\ \coeff_a^2 \\ - 2 \coeff_a \coeff_b \\ 0
  \end{pmatrix}^T
      \vect{M}
  =  (\omega_a-\omega_b)
  \begin{pmatrix}
    0 \\ 0 \\ 0 \\ 2 \coeff_a \coeff_b
  \end{pmatrix}.
\end{equation}
One may however immediately see this does not hold for the solution Eq.~(\ref{eq:13}) of our time-local master equation,
unless $K_a=K_b$.  We next find an alternative time-local equation of motion
for the observables $F_{ij}(t)$ that both satisfies this sum rule, and
gives the exact eigenvalues near degeneracy.

\subsection{Schr\"odinger picture Bloch-Redfield equation}
\label{sec:schr-pict-bloch-1}

The basis of the alternate approach is to consider the Born
approximation for the equation of motion, before making any Markov
approximation. We therefore first recall the form of the integro-differential equation for the density matrix after the Born approximation. In the interaction
picture this has the general form:
\begin{displaymath}
  \partial_t \rho^{(I)}(t)
  =
  \sum_{kl} \int^t dt^\prime 
  \eta_{kl}(t-t^\prime) [\hat{O}_k(t), [\hat{O}_l(t^\prime), \rho^{(I)}(t^\prime)]]
\end{displaymath}
where $\hat{O}_k(t)$ is an operator in the interaction picture and
$\eta_{kl}(\tau)$ accounts for the system-bath coupling, and the
integral over the bath density of states. From this one may derive the
population equation
\begin{displaymath}
  \partial_t F_{ij} = 
  \sum_{kl} \int^t_{-\infty} \!\!\!\!dt^\prime 
  \eta_{kl}(t-t^\prime)
  \left<  \left[ 
      \left[ \field^\dagger_i(t) \field^{}_j(t), \hat{O}_k(t) \right], 
      \hat{O}_l(t^\prime) 
    \right]
  \right>_I
\end{displaymath}
where $\langle \ldots \rangle_I = \text{Tr}[\ldots
\rho^{(I)}(t^\prime)]$.  The BR population equation then follows by
assuming $\rho^{(I)}(t^\prime)$ has a slow time dependence, and
performing the integral over $dt^\prime$ accounting only for the time
dependence of the interaction picture operator $\hat{O}_l(t^\prime)$.

If we focus on late times this procedure is somewhat strange, as it
is clear that for a problem which has a time-independent Hamiltonian
in the Schr\"odinger picture it is the density matrix in the
Schr\"odinger picture which will be time independent.  As such, an
alternate procedure suggests itself: to consider $\rho^{(I)}(t^\prime)
= e^{i \hat{H}_0 t^\prime} \rho^{(S)} e^{- i \hat{H}_0 t^\prime}$.
The explicit time dependence of this density matrix can be eliminated
using $\text{Tr}[ \hat{O} \rho^{(I)}(t^\prime)] = \text{Tr}\left[e^{- i
    \hat{H}_0 t^\prime} \hat{O} e^{i \hat{H}_0 t^\prime}
  \rho^{(S)}\right]$ so that we have:
\begin{displaymath}
  \partial_t F_{ij} = 
  \sum_{kl} \int_0^\infty \!\! d\tau
  \eta_{kl}(\tau)
  \left<  \left[ 
      \left[ \field^\dagger_i(\tau) \field^{}_j(\tau), \hat{O}_k(\tau) \right], 
      \hat{O}_l
    \right]
  \right>_S
\end{displaymath}
where  $\langle \ldots \rangle_S = \text{Tr}[\ldots \rho^{(S)}]$
and we have written $\tau=t-t^\prime$.
Following this prescription, one can again find an
equation for the vector of real quantities $\vect{f}$ 
in the form $\partial_t \vect{f} = -\vect{M}^S \vect{f} + \vect{f}_0$,
but the matrix $\vect{M}^S$ has a
different form.  The matrix is now given by:
\begin{equation}
\label{eq:17}
  \vect{M}^S
  =
  \begin{pmatrix}
    2\coeff_a^2 K^{\prime}_a  & 0 &
    \coeff_a \coeff^{}_b K^{\prime}_a &
    \coeff_a \coeff^{}_b K^{\prime\prime}_a
    \\
    0 & 2\coeff_b^2 K^{\prime}_b  &
    \coeff_b \coeff^{}_a K^{\prime}_b &
    \coeff_b \coeff^{}_a K^{\prime\prime}_b
    \\
    2 \coeff_a \coeff^{}_b K^{\prime}_a
    &
    2 \coeff_b \coeff^{}_a K^{\prime}_b
    &  \Gamma^S_0 & -E^S_0 
    \\
    -2 \coeff_a \coeff^{}_b K^{\prime\prime}_a
    &
    2 \coeff_b \coeff^{}_a K^{\prime\prime}_b
    & E^S_0 &  \Gamma^S_0 
  \end{pmatrix},
\end{equation}
where now $E^S_0 = (\omega_b-\coeff_b^2 K^{\prime\prime}_a) -
(\omega_a - \coeff_a^2 K^{\prime\prime}_b)$, and
$\Gamma^S_0=\coeff_a^2 K^{\prime}_b + \coeff_b^2 K^{\prime}_a$.  For
want of a better name, we refer to this as the Schr\"odinger picture
Bloch-Redfield (SpBR) equation.  The constant vector $\vect{f}_0$ is
unchanged.

The difference between the BR and SpBR equations has a simple
structure: it corresponds to swapping which frequency the bath is to
be sampled at in the third and fourth column. The origin of this
change is the unitary transformation $e^{i \hat{H}_0 t^\prime}$
between the interaction and Schr\"odinger pictures, which has the effect
of swapping time dependence of some ``off-diagonal'' terms.  These
small changes to the matrix $\vect{M}$ have several remarkable consequences.
Firstly we may immediately check that the sum rule as written in
Eq.~(\ref{eq:19}) is now exactly satisfied.  Secondly, one may
also consider the behavior of the eigenvalues of Eq.~(\ref{eq:17}).
Unlike Eq.~(\ref{eq:13}), there is no simple closed-form expression
for the eigenvalues in the general case --- Eq.~(\ref{eq:13}) was
special in having a structure that the secular equation could be
written as a quadratic in $\mu_i - \Re[\tilde{K}_a + \tilde{K}_b]$,
but this does not hold for Eq.~(\ref{eq:17}).  However, one can
perform perturbation theory around the point $\Delta=0$.  Clearly
the eigenvalues of $\vect{M}$ and $\vect{M}^S$ match at this point,
as the only distinctions occur if $K_a \neq K_b$.  Thus, using standard
(non-self-adjoint) perturbation theory in terms of the small parameters
$\Delta$ and $K_a-K_b$ one finds the lowest SpBR eigenvalue takes the form:
\begin{equation}
  \mu^S_0 
  =
  \frac{8\coeff_a^2 \coeff_b^2 \Delta^2 \bar{K}^\prime}{(\coeff_a^2 + \coeff_b^2)^3 |\bar{K}|^2}
  + \mathcal{O}(\Delta^3).
\end{equation}
Remarkably, this is identical to the exact solution, further
confirming the idea that this SpBR equation is an improvement over the
BR population equations discussed previously.

\begin{figure}[htpb]
  \centering
  \includegraphics[width=3.2in]{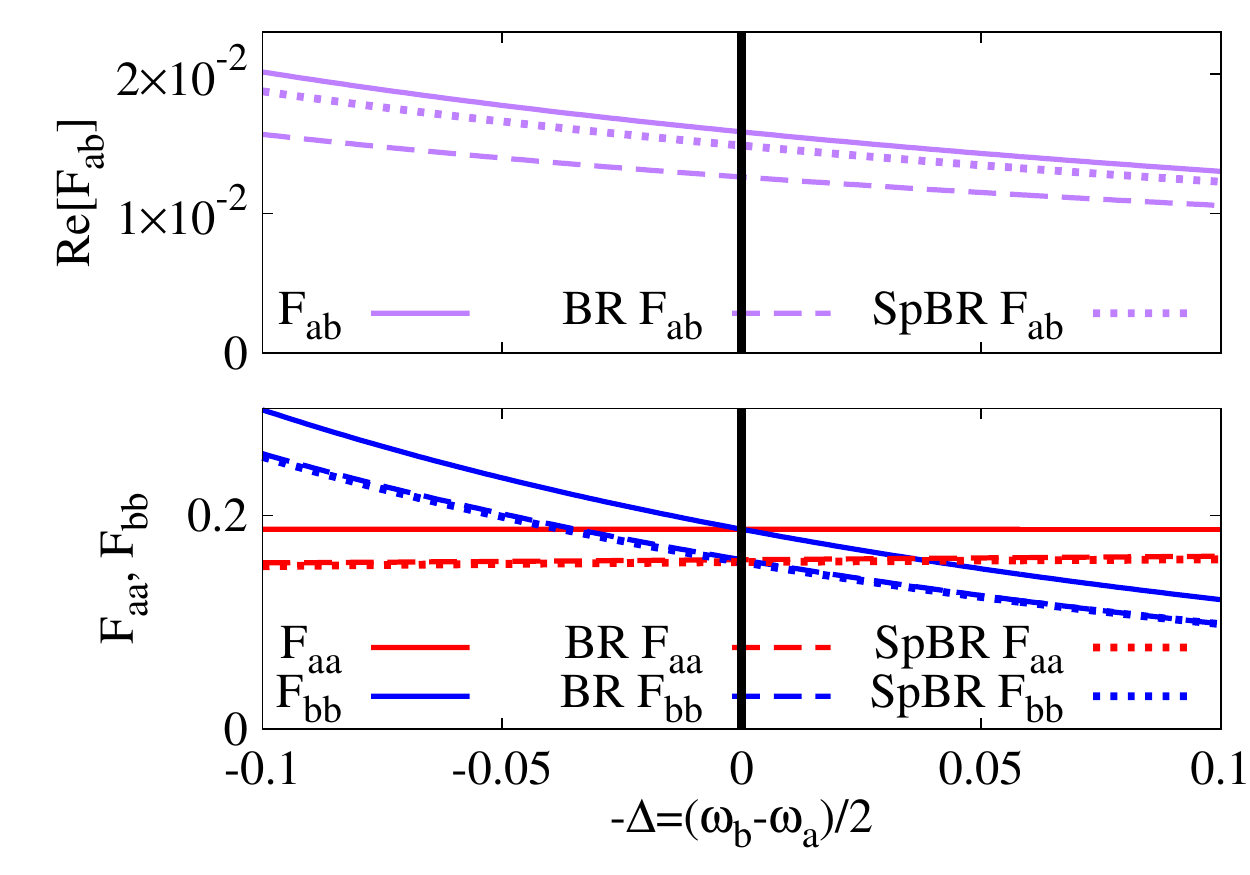}
  \caption{(Color online) Comparison of steady state values of $F_{ij}$ between
    the exact (solid), BR master equation (dashed) and SpBR
    master equation (dotted), plotted for a larger bath
    density of states $J_0=0.02$ and all other parameters as
    for Fig.~\ref{fig:coherence}.}
  \label{fig:ss}
\end{figure}

As we have already seen above, the BR master equation matches the
exact solution well as long as damping is weak enough and the Markov
approximation is well justified.  The decay rates near resonance have
further shown that while the BR master equation is correct to leading
order in the damping rate, the SpBR equation is correct to higher
order.  This suggests that as the damping rate becomes larger, the
SpBR may give a better numerical agreement with the exact solution.
This is indeed the case, and is shown in Fig.~\ref{fig:ss} where we
compare the steady state values of $F_{ij}$. 
Comparing the  coherence $F_{ab}(t)$, it is clear the SpBR matches
the exact solution better than the BR approach.  The lower panel
shows that the two theories give very similar results for the
populations.  For the
parameters corresponding to Fig.~\ref{fig:coherence} the BR and SpBR
lines would be indistinguishable.  

Note that at $\Delta=0$, the SpBR and BR formalisms are identical, and
so one might expect the results to match at this point.  However, the
matrices $\vect{M}$ are singular at $\Delta=0$ (as seen earlier from
their eigenvalues).  As such, the finite population and coherence at
$\Delta \to 0$ correspond to a singular limit.

\section{Extension to multiple baths}
\label{sec:extens-mult-baths}

Extending either the exact solution or the BR master equation to
multiple baths is simple. For the BR master equation, one just finds
a separate set of Lamb-shift terms $h_{ij}$ and dissipator terms
$L^\sigma_{ij}$ for each bath, so that Eq.~(\ref{eq:10}) involves a
summation over contributions from the baths.  Similarly, the
expressions for the matrix $\vect{M}$ follow as before, but now with a
sum over baths, and even the analytic form of the eigenvalues remains
true, with $\tilde{K}_i \mapsto \sum_n \tilde{K}^{(n)}_i$ in
Eq.~(\ref{eq:14}).  The exact solution is however more complicated.
Equation~(\ref{eq:2}) still holds, however there is a sum over baths,
and each term now acquires a bath label: $J(\nu), n_B(\nu),
W^{}_i(\nu,t) \mapsto J^{(n)}(\nu), n_{B}^{(n)}(\nu),
W^{(n)}_{i}(\nu,t)$.  The last of these quantities now has a more
complicated form
\begin{equation}
  \label{eq:18}
  W^{(n)}_i(\nu,t) =
  \int \frac{d\zeta}{2\pi} \frac{e^{-i\zeta t}}{\nu-\zeta-i0} 
  \sum_j
  \mathcal{G}_{ij}(\zeta)
  \coeff^{(n)}_j
\end{equation}\
where the matrix $\mathcal{G}$ can be defined in terms of its inverse,
$[\mathcal{G}(\zeta)^{-1}]_{ij} = i \delta_{ij} (\omega_i - \zeta - i0) +
\sum_n \coeff_i^{(n)\ast} \coeff_j^{(n)} K^{(n)\ast}(\zeta)$.  

In the presence of multiple baths, the singular behavior at
$\omega_a=\omega_b$ no longer occurs --- one may check this by
calculating the zeros of $\text{Det}[\mathcal{G}(\zeta)^{-1}]$: one now
finds there is no longer a zero mode, unless the coefficients
$\coeff_i^{(n)}$ happen to be parallel for different $n$.  The
physical origin of this is that with multiple linearly independent
baths there is no longer a linear combination of fields $\field_i$
which decouples from the baths, and so all modes are damped.  As such,
the collective dephasing model is never correct for predicting the
steady state coherence.  The non-secularized BR approach continues
  to correctly describe the system as one varies detuning.


\section{Conclusions}
\label{sec:conclusions}

In conclusion we have compared the exact and Bloch-Redfield solutions
for a system of two bosonic modes coupled to a common bath.  The
late-time behaviors show singular dependence on detuning: exactly on
resonance, significant coherence exists at late times, but for
arbitrarily small detuning the coherence drops to a smaller value
which depends on the frequency dependence of the density of states.
This singular limit appears only at late times, corresponding to a
slow decay rate for coherence that vanishes at the degenerate point.
All aspects of this behavior are reproduced correctly by a
non-secularized Bloch-Redfield theory, whereas secularization leads to
incorrect predictions.  The Bloch-Redfield theory does not guarantee
positivity, nonetheless one can prove that the equations describe
bounded dynamics of physical observables, as long as the Markov
approximation remains valid. A modification to the Bloch-Redfield
theory --- assuming it is the Schr\"odinger picture density matrix that
evolves slowly, rather than the interaction picture one --- leads to
an improved time-local theory which satisfies required sum rules and
exactly matches damping rates near resonance.

\acknowledgments{We are very happy to acknowledge discussions with
  V. Oganasyan and T. Stace.  PGK and JK acknowledge financial support from EPSRC
  program ``TOPNES'' (EP/I031014/1).
  PGK acknowledges support from EPSRC (EP/M010910/1).  
  JK acknowledges support from the Leverhulme Trust (IAF-2014-025) and
  EPSRC program ``Hybrid-Polaritonics'' (EP/M025330/1).
  HMC acknowledges support from EPSRC (EP/G03673X/1).
  BWL acknowledges support from the Leverhulme Trust (RPG-080), and the joint EPSRC (EP/I035536) / NSF (DMR-1107606) Materials World Network grant.}


%

\end{document}